\documentclass[prd,aps,showpacs,preprintnumbers]{revtex4}
\usepackage{amssymb}
\usepackage{amsmath}
\usepackage{graphicx}
\usepackage{dcolumn}
\usepackage{bm}

\input{tcilatex}
\begin{document}

\title{ Nucleosynthesis at Finite Temperature and Density}
\author{Samina S. \textsc{Masood}}
\email{masood@uhcl.edu}
\affiliation{Department of Physics, University of Houston Clear Lake, Houston, TX 77058}

\begin{abstract}
We study the finite temperature and density effects on beta decay rates to
compute their contributions to nucleosynthesis in the early universe and
compact stars. We express nucleosynthesis parameters as a function of
temperature and density in different astronomical systems of interest. It is
explicitly shown that the chemical potential in the core of supermassive and
superdense stars affect beta decay and their helium abundance but the
background contributions is still dependent on relative temperature. We
calculate this contribution for $T\ll m\ll \mu .$It has been noticed that
the acceptable background contribution are obtained for comparatively larger
values of T as temperature plays a role of regulating parameter in an
extremely dense system.
\end{abstract}

\maketitle

\section{Introduction}

Standard big bang model (SBBM) of the universe [1] indicates that the
universe went through nucleosynthesis when it was cooled to $10^{10}$ K.
Beta decay processes started when the baryon density $\eta _{B}$ was as low
as $10^{-10}$; Big bang nucleosynthesis (\ BBN)\ in the universe started
around the same time. It is known that the beta decay rates [1,2] depend on
masses of particles and the phase space. Field theory of quantum
electrodynamics (QED) at finite temperature and density (FTD) shows that the
electron mass, wavefunction and charge are modified in a hot and dense
medium. Change in the physical properties of electrons is determined through
their interaction with particles in a hot and dense medium while they
propagate through the medium. These physically measureable values of
parameters correspond to the effective parameters of the theory in that
medium.

SBBM of the universe, the most well-known model of cosmology, predicts the
abundances of light elements in the early universe. Beta decay processes are
observed on Earth as well as in the extremely hot universe and inside the
extremely hot and dense stellar cores, whereas the nucleosynthesis only
occurs under the special conditions of temperatures and densities such as in
the primordial universe and in the stellar cores. Since beta decay is a weak
process and nucleosynthesis is expected to start after the decoupling
temperature [3,4], QED is not enough to fully describe nucleosynthesis. One
can think about a major role of electroweak theory at FTD in
nucleosynthesis. However, the concentrations of hot and dense W's and Z's in
the background are totally insignificant as compared to photons. The
electroweak corrections of the background are suppressed by the heavy masses
of electroweak mediators W and Z. Neutrino mass being tiny enough (even if
it exists in the extended standard models [3-5] ) is ignored for all
practical purposes. However, some of the corrections due to the form factors
of neutrinos [6-10] cannot be ignored at these temperatures and densities.
Although, for extremely high temperatures at the beginning of the universe,
we may not be able to totally ignore the electroweak background corrections
[11] for the study of leptogenesis, but at that point, extensions of
standard models have to be used (See for example: [12-13]).

Beta decay processes are studied in a hot and dense media to accurately
calculate their contributions to nucleosynthesis. However, all
nucleosynthesis temperatures are well below the electroweak scale and we can
easily ignore thermal contributions to electroweak processes in this range.
If the neutrino is not massless and we use minimal extension of the standard
model to work with the tiny mass of Dirac type neutrino, the weak processes
have to be incorporated. Also the properties of neutrinos are significantly
modified in hot and dense media at this scale, provided the neutrinos have
non-zero mass due to extensions in the standard model, we would have to
include thermal contributions due to the non-zero mass of neutrino in the
minimal extensions ( or other extensions) of the standard model. However, in
this paper we just restrict ourself to the standard electroweak model with
the massless neutrino and exclusively study the QED type FTD\ corrections
[14-26]only. For this purpose, we consider previously calculated
relationships of electron mass with nucleosynthesis parameters, such as beta
decay rate and helium abundance in the early universe [1,2].

High energy physics provides a theoretical justification of the SBBM. When
the universe was less than a second old, it was extremely hot and
electron-positron pairs were created as the first matter particles.
Properties of electrons in the very early environment of the universe were
not the same as they are in a vacuum. The behavior of electrons in the very
early universe can be predicted incorporating thermal background effects on
the physical properties of electrons. We use the renormalization scheme of
QED to determine the renormalization constants of QED, such as electron
selfmass, charge and wavefunction, to study the physical properties of
electrons at high temperatures.

Renormalization of QED at finite temperature and density ensures a
divergence free QED in hot and dense media. The most general calculations of
the first order thermal loop corrections to electron selfmass, charge and
wavefunction are performed in detail, incorporating the background density
effects through the chemical potential [23-26]. Calculations of the second
order corrections, at finite temperatures [21,22], to the renormalization
constants, in different ranges of temperatures are already there in
literature. However, it is not possible to separate thermal corrections from
the vacuum contributions. Also the validity of the renormalization scheme
fully justifies that the second order corrections are significantly small as
compared to the first order thermal corrections.

In the next section we briefly mention the calculational scheme and rewrite
some of the selfmass of electron expressions in more useful form for the
more relevant regions of temperature and chemical potentials of
astrophysical systems.

Section 3 is devoted to a discussion of nucleosynthesis at finite
temperature and density corrections to the electron mass in the early
universe and for the stellar cores. Just for simplicity, we have not
included the effect of strong magnetic fields in the core of neutron stars
as it has to be studied separately, in detail, because of the complexity of
the issue. Also, interestingly, due to the recently observed existence [27]
of superfluids, a comprehensive study of this aspect of the problem is
demanded. High abundance of helium at high density and low temperature, in
the presence of strong magnetic fields may provide favorable conditions for
superfluids.

\section{\protect \bigskip Calculational Scheme}

We summarize the previously calculated results of the renormalization
constants of QED in the real-time formalism, up to the one loop level, at
FTD. It is possible to separate out the temperature dependent contributions
from the vacuum contributions as the statistical distribution functions
contribute additional statistical terms both to fermion and boson
propagators in the form of the Fermi-Dirac distribution and the
Bose-Einstein distribution functions, respectively. The Feynman rules of
vacuum theory are used with the statistically corrected propagators[15-17]
given as

\begin{equation*}
D(k)=\frac{i}{k^{2}}+2\pi n_{B}(k_{0})\delta (k^{2}),
\end{equation*}%
for bosons, and \bigskip 
\begin{equation}
S_{F}(p)=\frac{i}{\NEG{p}-m+i\varepsilon }-2\pi \delta (p^{2}-m^{2})[\theta
(p_{0})n_{F}^{+}(p,\mu )+\theta (-p_{0})n_{F}^{-}(p,\mu )]\ ,
\end{equation}%
for fermions, where the corresponding energies of electrons (positrons) are
defined as

\begin{equation*}
E_{p,k}=\sqrt{(\overrightarrow{p},\overrightarrow{k})^{2}-m^{2}}.
\end{equation*}%
$\mu $ is the chemical potential which is assigned a positive sign for
particles and negative sign for antiparticles. Fermion distribution function
at FTD can then be written as%
\begin{equation}
n_{F}^{\pm }(p,\mu )\equiv n_{F}(p\pm \mu )=\frac{1}{e^{-\beta (\mid
E_{p}\mid \pm \mu )}+1},
\end{equation}%
where the positive sign corresponds to electron and negative to positron.
This sign difference ( in $n_{F}^{\pm }(p,\mu )$) determines the difference
in behavior of particles (antiparticles) in a dense background.

It is obvious from Eqs. (1) and (2) that the photons and electrons
propagated differently, in the beginning of the hot early universe, right
after its creation. Photons, being massless particles, exhibit zero chemical
potential and no density effects, whereas the electrons (positrons)
propagation in the medium help to understand several issues in dense media
which are out of scope of this paper.

The electron mass, wavefunction and charge are then calculated in a
statistical medium using Feynman rules of QED, with the modified propagators
given in Eqs. (1) and (2). These renormalization constants are evaluated for
different hot and dense systems to understand the propagation of electrons
in such media. These renormalization constants behave as effective
parameters of hot and dense systems. We briefly overview the calculations of
the relevant parameters of QED at FTD and explain how the renormalization
constants can be used to describe the physical behavior of the hot and dense
systems.

\subsection{Selfmass of Electron\protect \bigskip}

The renormalized mass of electrons $m_{R}$\ can be represented as a physical
mass $m_{phys}$ of electron and is defined in a hot and dense medium as,

\begin{equation}
m_{R}\equiv m_{phys}=m+\delta m(T=0)+\delta m(T,\mu ).
\end{equation}%
where $m$ is the electrons mass at zero temperature; $\delta m(T=0)$\
represents the radiative corrections from vacuum and $\delta m(T,\mu )$ is
the perturbative corrections to the mass (selfmass) due to its interaction
to the statistical background at FTD. Thermal effects are computed by means
of the particle interaction with the hot particles of the medium at
temperature T; the density effects are estimated in terms of the chemical
potential $\mu $. The physical mass can get radiative corrections at
different orders of $\alpha $ and can be written as:

\begin{equation}
m_{phys}\cong m+\delta m^{(1)}+\delta m^{(2)},
\end{equation}%
where $\delta m^{(1)}$ and $\delta m^{(2)}$\ are the shifts in the electron
mass in the first and second order in $\alpha $, respectively. This
perturbative series can go to any order of $\alpha $, as long as it is
convergent. The physical mass is calculated by locating the pole of the
fermion propagator $\frac{i(\NEG{p}+m)}{p^{2}-m^{2}+i\varepsilon }$ in
thermal background.$\ $For this purpose, we sum over all the same order
diagrams at FTD. Renormalization is established by demonstrating the
order-by-order cancellation of singularities at finite temperatures and
densities. All the terms from the same order in $\alpha $ are combined
together to evaluate the same order contribution to the physical mass given
in Eq.(4) and are required to be finite to ensure order by order
cancellation of singularities. The physical mass in thermal background up to
order $\alpha ^{2}$ [20-21] is calculated at finite temperature, using the
renormalization techniques of QED. Higher order background contributions to
electron mass, due to the chemical potential are still to be computed.
However, following the renormalization scheme of vacuum, we may write the
selfmass term as

\begin{equation}
\Sigma (p)=A(p)E\gamma _{_{0}}-B(p)\vec{p}.\vec{\gamma}-C(p),
\end{equation}%
where $A(p)$, $B(p)$, and $C(p)$ are the relevant coefficients and are
modified at FTD. Taking the inverse of the propagator where the momentum and
mass terms separated as,

\begin{equation}
S^{-1}(p)=(1-A)E\gamma ^{o}-(1-B)p.\gamma -(m-C).
\end{equation}

The temperature-dependent radiative corrections to the electron mass are
obtained from the FTD propagators in Eq. (4). These corrections are
rewritten in terms of the boson loop integral I's and the fermion loop
integrals J's with the one loop level as,%
\begin{eqnarray}
E^{2}-|\mathbf{p}|^{2} &=&m^{2}+\frac{\alpha }{2\pi ^{2}}%
(I.p+J_{B}.p+m^{2}J_{A})  \notag \\
&\equiv &m_{phys}^{2},
\end{eqnarray}

\bigskip with

\begin{eqnarray}
I_{A} &=&8\pi \int \frac{dk}{k}n_{B}(k).  \notag \\
I^{\mu } &=&2\int \frac{d^{3}k}{k}n_{B}(k)\frac{k^{\mu }}{p_{\nu }k^{\nu }}%
=2\int \frac{d^{3}k}{k}n_{B}(k)\frac{(k_{0,}\overrightarrow{k})}{E_{p}k_{0}-%
\overrightarrow{p}.\overrightarrow{k}},  \notag \\
J_{A} &=&\int \frac{d^{3}l}{E_{l}}n_{F}(E_{l},\pm \mu )\left[ \frac{1}{%
E_{p}E_{\ell }+m^{2}-\overrightarrow{p}.\overrightarrow{l}}-\frac{1}{%
E_{p}E_{\ell }-m^{2}+\overrightarrow{p}.\overrightarrow{l}}\right] , \\
J_{B}^{\mu } &=&\int \frac{d^{3}l}{E_{l}}n_{F}(E_{l},\pm \mu )\left[ \frac{%
(E_{p}+E_{\ell },\overrightarrow{p}+\overrightarrow{l})}{E_{p}E_{l}+m^{2}-%
\overrightarrow{p}.\overrightarrow{l}}-\frac{(E_{p}-E_{l},\overrightarrow{p}+%
\overrightarrow{l})}{E_{p}E_{l}-m^{2}+\overrightarrow{p}.\overrightarrow{l}}%
\right] .\text{\ }  \notag
\end{eqnarray}

\qquad \qquad \qquad \qquad \qquad \qquad \qquad \qquad \qquad \qquad \qquad
\qquad \qquad \qquad \qquad \qquad \qquad \qquad \qquad \qquad \qquad 
\begin{equation}
I.p=\frac{4\pi ^{3}T^{2}}{3},
\end{equation}%
and \qquad \qquad \qquad \qquad \qquad \qquad \qquad \qquad \qquad \qquad
\qquad \qquad \qquad \qquad \qquad \qquad \qquad \qquad 
\begin{eqnarray}
J_{B}.p &=&8\pi \lbrack I_{1}(m\beta ,\pm \mu )-\frac{m^{2}}{2}I_{2}(m\beta
,\pm \mu )]  \notag \\
&=&8\pi \left[ \frac{m}{\beta }a(m\beta ,\pm \mu )-\frac{m^{2}}{2}b(m\beta
,\pm \mu )-\frac{1}{\beta ^{2}}c(m\beta ,\pm \mu )\right] .
\end{eqnarray}%
Thus up to the first order in $\alpha ,$ FTD corrections to the electron
mass at $\mu <T$ can be obtained as%
\begin{equation}
m_{phys}^{2}=m^{2}\left[ 1-\frac{6\alpha }{\pi }b(m\beta ,\pm \mu )\right] +%
\frac{4\alpha }{\pi }\left[ mT\text{ }a(m\beta ,\pm \mu )+\frac{2}{3}\alpha
\pi T^{2}-\frac{6}{\pi ^{2}}c(m\beta ,\pm \mu )\right] .
\end{equation}%
\begin{eqnarray}
\frac{\delta m^{1}}{m} &\simeq &\frac{1}{2m^{2}}\left(
m_{phys}^{2}-m^{2}\right)  \notag \\
&\simeq &\frac{\alpha \pi T^{2}}{3m^{2}}\left[ 1-\frac{6}{\pi ^{2}}c(m\beta
,\pm \mu )\right] +\frac{2\alpha }{\pi }\frac{T}{m}a(m\beta ,\pm \mu )-\frac{%
3\alpha }{\pi }b(m\beta ,\pm \mu ).
\end{eqnarray}%
where $+\mu $ (-$\mu $) correspond to the chemical potential of electron
(positron) and correspond to the density of the system. $\frac{\delta m}{m}$
is the relative shift in electron (positron) mass due to finite temperature
and density of the medium, determined in Ref. [17] with%
\begin{eqnarray}
a(m\beta ,\pm \mu ) &=&\ln (1+e^{-\beta (m\pm \mu )}),  \notag \\
b(m\beta ,\pm \mu ) &=&\dsum \limits_{n=1}^{\infty }(-1)^{n}e^{\mp n\beta
\mu }\func{Ei}(-nm\beta ), \\
c(m\beta ,\pm \mu ) &=&\dsum \limits_{n=1}^{\infty }(-1)^{n}\frac{e^{-n\beta
\left( m\pm \mu \right) }}{n^{2}}.  \notag
\end{eqnarray}%
The validity of Eq.(13) can be ensured for T\TEXTsymbol{<}2MeV in the early
universe where $\left \vert \mu \right \vert $ is ignorable (see Ref.[7],
for details). The big bang theory of cosmology and all the observational
data of the universe agree that primordial nucleosynthesis occured when the
universe cooled down to around $10^{10}$K. On the other hand,
renormalization of quantum electrodynamics (QED) in hot and dense media
indicate that the hot medium contribution at $T=m$ is different for a
heating and a cooling system [3,4]. This disagreement supports the big bang
model as it indicates that the universe was going through nucleosynthesis at
those temperatures and the compositional change in the universe.

The convergence of Eq.(4) can be ensured [3,4] at $T\leq 2MeV$ as $\frac{%
\delta m}{m}$ is sufficiently smaller than unity within this limit[15]. This
scheme of calculations will not work for higher temperatures and the first
order corrections may exceed the original values of QED parameters, after $%
2MeV$. At low temperature $T<m$, the functions $a(m\beta ,\pm \mu )$, $%
b(m\beta ,\pm \mu )$, and $c(m\beta ,\pm \mu )${\LARGE \ }fall off in powers
of $e^{-m\beta }$ in comparison with $\left( \frac{T}{m}\right) ^{2}$ when ($%
\mu <m<T$ ) and can be neglected in the low temperature limit giving,%
\begin{equation}
\frac{\delta m}{m}\overset{T\ll m}{\rightarrow }\frac{\alpha \pi T^{2}}{%
3m^{2}}.
\end{equation}%
In the high-temperature limit, neglecting $\mu $, $a(m\beta ,\pm \mu )$ and $%
b(m\beta ,\pm \mu )$ are still vanishingly small whereas $c(m\beta ,\pm \mu
)\longrightarrow -\pi ^{2}/12$, yield%
\begin{equation}
\frac{\delta m}{m}\overset{T\gg m}{\rightarrow }\frac{\alpha \pi T^{2}}{%
2m^{2}}.
\end{equation}%
The above equations give $\frac{\delta m}{m}=7.647\times 10^{-3}\frac{T^{2}}{%
m^{2}}$ for low temperature and $\frac{\delta m}{m}=1.147\times 10^{-2}\frac{%
T^{2}}{m^{2}}$ for high temperature, showing that the rate of change of mass$%
\frac{\delta m}{m}$ is larger at $T>m$ as compared to $T<m$. Subtracting
eq.(12) from (13), the change in $\frac{\delta m}{m}$ between low and high
temperature ranges can be written as

\begin{equation}
\Delta (\frac{\delta m}{m})=\pm \frac{\alpha \pi T^{2}}{6m^{2}}=\pm
3.8\times 10^{-3}\frac{T^{2}}{m^{2}}.
\end{equation}%
Nucleosynthesis is held responsible for that. Eqs. (14) and (15) show that
thermal corrections to the electron selfmass are expressed in terms of T/m
both for low T and for high T. It is only during the nucleosynthesis that
selfmass deviates from the T/m and has to be expressed in terms of a$_{i}$
functions derived by Masood [25, 26]. Actually since there is no significant
change in the density of electrons or photons, the thermal contributions to
the beta decay and other nucleosynthesis parameters, including the helium
yield, do not change much as long as the low T or high T approximations are
valid to use Eqs. (14,15). However at the higher loop level, the low T and
high T approximations are not so well described and they are tied up with
vacuum effects. Moreover, the selfmass expressions are much more complicated
to retrieve thermal corrections to helium yield and the above given
relations between the selfmass and the nucleosynthesis parameers have to be
revised as the phase space for beta processes changes at higher loops and
will involve Boltzman Equations. So we restrict ourselves to the one loop
corrections as they are the only relevant corrections at such temperatures.
However, we give a plot of Eqs. (14) and (15) in Figure 1 to explictely show
the disconnected region at T=m if we approach this point from Eq.(14) or
Eq.(15). This descripency appears because of the nucleosynthesis around $%
T\sim m$, although both Eqs.(14) and Eq.(15) have been derived from the same
master Eq. (12)

\FRAME{ftbpFU}{1.7331in}{1.3405in}{0pt}{\Qcb{Thermal mass of electrons is
plotted as a function of temperature. The low T behavior and high T values,
derived from the same expression, does not coincide at T=m, indicating the
presence of nucleosynthesis around those temperatures. }}{\Qlb{Figure 1}}{%
electronmass.eps}{\special{language "Scientific Word";type
"GRAPHIC";maintain-aspect-ratio TRUE;display "USEDEF";valid_file "F";width
1.7331in;height 1.3405in;depth 0pt;original-width 11.7796in;original-height
9.1022in;cropleft "0";croptop "1";cropright "1";cropbottom "0";filename
'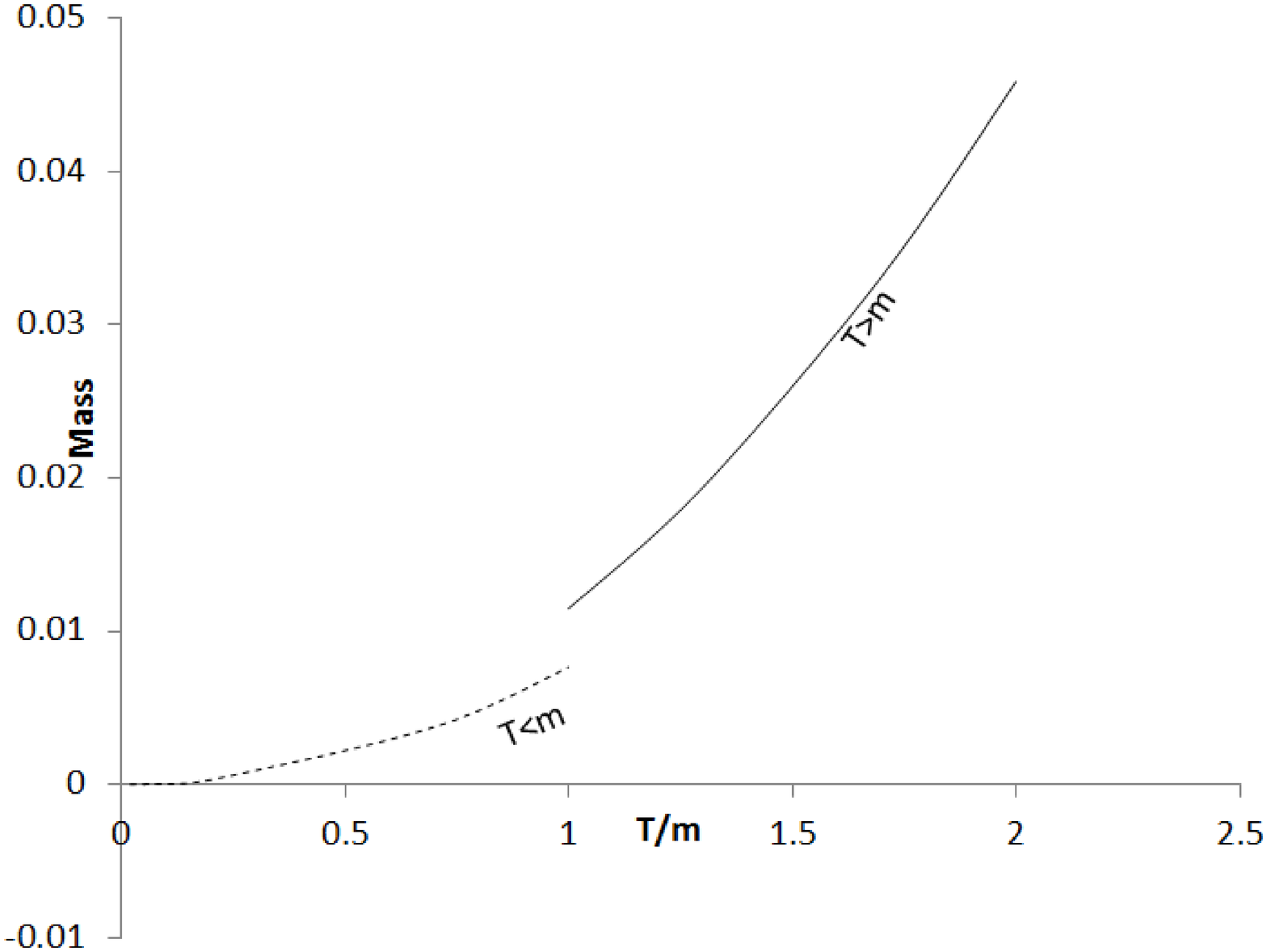';file-properties "XNPEU";}}

When the density effects are not ignorable, the perturbative series is still
valid at much higher temperatures, due to the reason that the growth of mass
is slowed down significantly in a dense system. High densities and smaller
mean free paths automatically ensure the validity of the perturbative
expansion as the argument of the exponential changes from $\beta m$ to $%
\beta (m\pm \mu )$ and the expansion parameter changes from $m/T$ to $(m\pm
\mu )/T$ as $T=1/\beta $, for such systems. When the chemical potential is
large, it can overcome thermal effects as we deal with $\beta \mu $ and not $%
\beta m$, the expansion parameters. Especially for electrons ( at $\mu \gg
m\geq T$), the distribution function $n_{F}^{\pm }(p,\mu )$ reduces to the
function $\theta (\mu -E_{l})$ for electrons, (and vanishes for positrons)
providing $\mu $ as an upper limit to $E_{l}$, such that the integration is
simplified as

\begin{eqnarray}
\int_{m}^{\infty }\frac{dE_{l}}{E_{l}}\theta (\mu -E_{l}) &=&\ln \frac{\mu }{%
m},  \notag \\
\int_{m}^{\infty }E_{l}dE_{l}\theta (\mu -E_{l}) &=&\frac{1}{2}\left( \mu
^{2}-m^{2}\right) , \\
\int_{m}^{\infty }\frac{dE_{l}}{E_{l}^{3}}\theta (\mu -E_{l}) &=&-\frac{1}{2}%
\left( \frac{1}{\mu ^{2}}-\frac{1}{m^{2}}\right) .  \notag
\end{eqnarray}

Inserting the results of integrations of Eq. (17) into Eqs. (8) and (9), we
obtain

\begin{center}
\begin{eqnarray}
J_{A} &\simeq &-8\pi \ln \frac{\mu }{m}+8\pi \frac{E_{p}^{2}}{\mu ^{2}}%
\left( 1-\frac{\mu ^{2}}{m^{2}}\right) ,  \notag \\
\int_{0}^{\infty }\frac{l^{2}dl}{E_{l}}\theta (\mu -E_{l}) &=&-\frac{m^{2}}{2%
}\ln \frac{\mu }{m}-\frac{m^{2}}{2}\left( 1-\frac{\mu ^{2}}{m^{2}}\right)
\left( 1-\frac{m^{2}}{\mu ^{2}}\right) \\
\frac{J_{B}^{0}}{E} &\simeq &\pi \left( 1-\frac{\mu ^{2}}{m^{2}}\right)
\left( \frac{2m^{2}}{pE}\ln \frac{1-v}{1+v}-\frac{E^{2}}{4m^{2}}\right)
-4\pi \ln \frac{\mu }{m},  \notag
\end{eqnarray}
\end{center}

giving

\begin{equation}
m_{phys}^{2}\simeq m^{2}-\frac{6\alpha }{\pi }\ln \frac{\mu }{m}+\frac{%
2\alpha }{\pi }m^{2}\left( 1-\frac{\mu ^{2}}{m^{2}}+\frac{2p^{2}}{\mu ^{2}}%
-1\right) ,
\end{equation}

and

\begin{equation}
\frac{\delta m}{m}(T,\mu )\simeq -\frac{3\alpha }{\pi }\ln \frac{\mu }{m}+%
\frac{\alpha }{\pi }\left( 1-\frac{\mu ^{2}}{m^{2}}\right) \left( 3\frac{%
m^{2}}{\mu ^{2}}+\frac{2p^{2}}{\mu ^{2}}-1\right) .
\end{equation}

Eqs. (19) and (20) give the electron selfmass for the extremely dense
stellar cores which have very high temperatures, but due to the extremely
dense situation, cannot be treated as purely hot systems. Neutron stars
provide a good example of such systems. However, in neutron stars, high
magnetic field effects are not ignorable either, though they are out of the
scope of this paper. Eq. (20) shows that the extremely large values of $\mu $
will lead to the dominent behavior of electron mass as

\begin{equation}
\frac{\delta m}{m}(T,\mu )\simeq \frac{\alpha }{\pi }\frac{\mu ^{2}}{m^{2}}
\end{equation}

Eq. (21) shows the mass dependence on chemical potential is just as it were
at T, in the extremly large chemical potential values and selfmass of
electron grows as $\frac{\mu ^{2}}{m^{2}}$ and can be plotted as in Figure 2.

\FRAME{ftbpFU}{1.8974in}{1.3197in}{0pt}{\Qcb{Electron selfmass as a function
of chemical potential. for $\protect \mu >T>m$}}{\Qlb{Figure }}{mumass.eps}{%
\special{language "Scientific Word";type "GRAPHIC";maintain-aspect-ratio
TRUE;display "USEDEF";valid_file "F";width 1.8974in;height 1.3197in;depth
0pt;original-width 8.4743in;original-height 5.8816in;cropleft "0";croptop
"1";cropright "1";cropbottom "0";filename '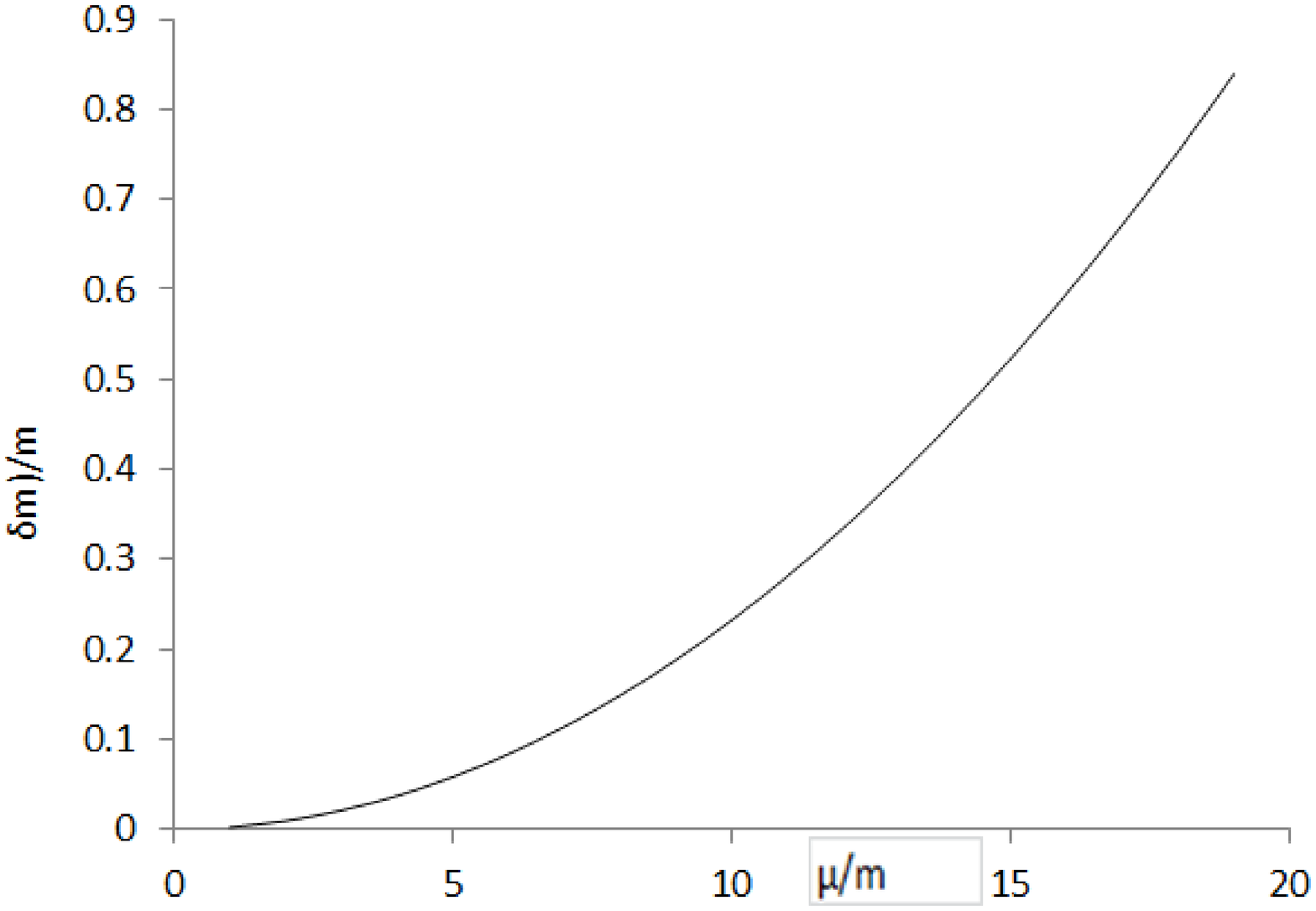';file-properties
"XNPEU";}}

Using the electron selfmass contribution from FTD\ background, we can
calculate the background contribution to the helium abundance parameter,
corresponding to the astrophysical systems of interest. This contribution is
small, still nonignorable.

\section{\protect \bigskip Beta Decay and Nucleosynthesis}

The big bang model of the early universe indicates that nucleosynthesis
takes place around the temperatures of the electron mass, that is around $%
10^{10}$ K. Nucleosynthesis was initiated by the creation of protons as
hydrogen nuclei. A proton can capture an electron to create neutrons which
can decay back to a proton through beta decay. Therefore, the beta decay
processes are usually studied in detail to understand the start of
nucleosythesis. The abundances of light elements are related to the neutron
to proton ratio as well as nucleon to photon ratio, calculated in the medium
at the time of nucleosynthesis. Beta-processes kept the ratio between
protons and neutrons in all the relevant channels and photons were regulated
by the background temperatures [1, 2, 28]. \ 

\begin{equation}
\Delta Y=0.2\frac{\Delta \tau }{\tau }=-0.2\frac{\Delta \lambda }{\lambda },
\end{equation}%
where $\frac{\Delta \tau }{\tau }$ is relative change in neutron half life
and $\frac{\Delta \lambda }{\lambda }$ is relative change in neutron decay
rate.

\subsection{\protect \bigskip First Order Contributions from Electron Selfmass%
}

Neutron decay rate or half life can easily be related to the electron mass.
It has been explicitly shown [1,2, 28] that the radiative corrections,
especially thermal background contributions to beta decay rate and all the
other parameters can be expressed in terms of the selfmass of electron. So
the radiative corrections to beta decay rate can be expressed as

\begin{equation}
\frac{\Delta \lambda }{\lambda }=-0.2\left( \frac{m}{T}\right) ^{2}\frac{%
\delta m}{m},
\end{equation}%
with $m$ as the mass of the propagating electron, $T$ is the temperature of
background heat bath and $\frac{\delta m}{m}$ is the radiative corrections
to electron mass due to its interaction with thermal medium. In the early
universe the temperature effects were dominant with ignorable density
effects as the chemical potential $\mu $\ of the particles satisfies the
condition $\mu /T\leqslant 10^{-9}$. The neutrino temperature $T_{\nu }$ can
be written as%
\begin{equation*}
\frac{\Delta T_{\nu }}{T_{\nu }}=-0.1\left( \frac{m}{T}\right) ^{2}\frac{%
\delta m}{m},
\end{equation*}%
\begin{equation*}
\frac{\Delta T_{\nu }}{T}=-\left( \frac{m}{5T}\right) ^{2}\frac{\delta m}{m}.
\end{equation*}

\bigskip Considering all of the three generation of leptons, we can express $%
\Delta Y$,

\begin{equation}
\Delta Y\simeq -0.01\frac{\Delta T_{\nu _{e}}}{T_{\nu _{e}}}+0.04\frac{%
\Delta T_{\nu _{\mu }}}{T_{\nu _{\mu }}}+0.04\frac{\Delta T_{\nu _{\tau }}}{%
T_{\nu _{\tau }}}.
\end{equation}

However, we will just study the contribution of the first term. The
contribution due to the $\nu _{\mu }$ and $\nu _{\tau }$ background will be
ignored because these neutrinos do not decouple until $T=3.5MeV$ and we are
working below those temperatures in the early universe.

\begin{equation}
\frac{\Delta \rho _{_{T}}}{\rho _{_{T}}}=-\left( \frac{m}{4T}\right) ^{2}%
\frac{\delta m}{m}.
\end{equation}

The total energy density $\rho _{_{T}}$\ of the universe affects the
expansion rate of the universe H%
\begin{equation}
H=\left( \frac{8}{3}\pi G\rho _{_{T}}\right) ^{1/2}
\end{equation}

giving

\begin{equation}
H=\left( \frac{8}{3}\pi G\rho _{_{T}}(1-\left( \frac{m}{4T}\right) ^{2}\frac{%
\delta m}{m})\right) ^{1/2}
\end{equation}

which corresponds to the change in H as

\begin{equation}
\frac{\Delta H}{H}\simeq -0.5\left( \frac{m}{4T}\right) ^{2}\frac{\delta m}{m%
}
\end{equation}

and Eq. (22) can be re-written as%
\begin{equation}
\Delta Y=0.04\left( \frac{m}{T}\right) ^{2}\frac{\delta m}{m}.
\end{equation}%
Thermal contribution to beta decay for $T\leq m$ is -0.00153 and at $T\geq m$
is -0.00229, whereas thesel contribution to helium synthesis parameter is
0.000306 and for large T it is 0.000459. It gives thermal corrections to Y
for a heating universe: 0.03\% and for a cooling universe it is 0.045 \% of
the accepted value of around 0.25.

\bigskip

\FRAME{ftbpFU}{1.9207in}{1.4857in}{0pt}{\Qcb{A comparison of helium
abundance at low temperature and the helium abundance at high temperature.}}{%
}{heliumabundance.eps}{\special{language "Scientific Word";type
"GRAPHIC";maintain-aspect-ratio TRUE;display "USEDEF";valid_file "F";width
1.9207in;height 1.4857in;depth 0pt;original-width 11.7796in;original-height
9.1022in;cropleft "0";croptop "1";cropright "1";cropbottom "0";filename
'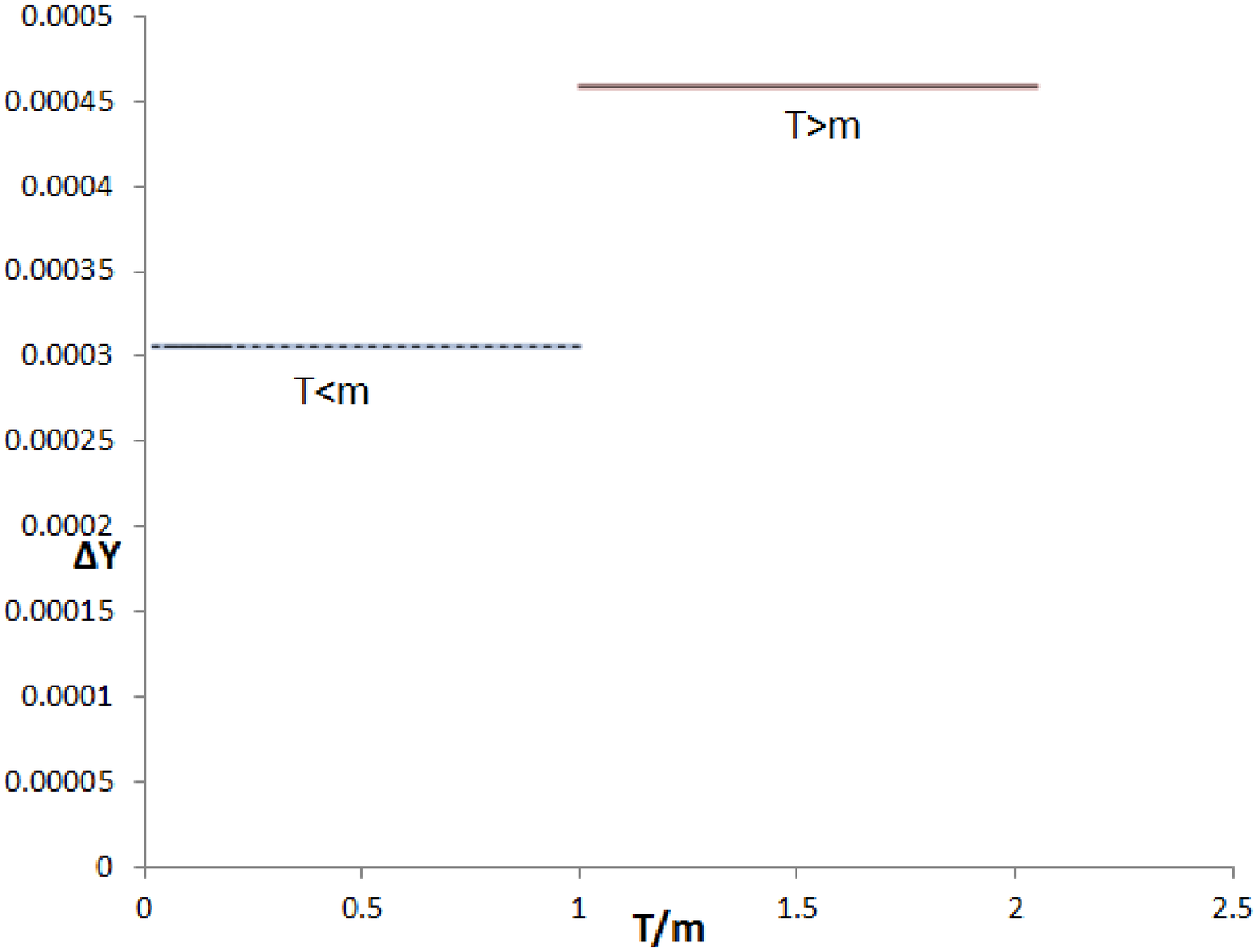';file-properties "XNPEU";}}

$T\sim $ $m$ range of temperature is particularly interesting \ from the
point of view of primordial nucleosynthesis. It has been found that some
parameters in the early universe such as the energy density and the helium
abundance parameter $Y$ become a slowly varying function of temperature [2];
whereas they remain constant before and after the nucleosynthesis as the
quadratic term in selfmass contribution from the background cancels out when
Eqs. (14) and (15) are substituted.

The chemical potential selfmass contributions can also be calculated from
Eq. (21) directly as

\begin{equation}
\Delta Y=0.04\left( \frac{m}{T}\right) ^{2}\frac{\alpha }{\pi }\frac{\mu ^{2}%
}{m^{2}}.
\end{equation}

for the chemical potential $\mu $ sufficiently greater than temperature T as
well as the electron mass. $T<m$ and $\mu >m$. However, it can be seen that
the presence of (m/T) factor in Eq. (30) ensures that the helium synthesis
will blow up at low temperature or the system will not maintain equilibrium
at low T. Therefore, temperature acts like a regulating parameter at high
chemical potential indicating that inside the stellar cores with large
chemical potential of electron, the temperature has to be high.

\subsection{\protect \bigskip Second Order Contributions from Electron
Selfmass}

Study of the early universe has passed through the stage of theoretical
prediction to observations and testing. COBE\ (\ Cosmic Background
Explorer), WMAP\ ( Wilkinsin Microwave Anisotropy Probe) and LHC (Large
Hadron Collider) provide data [29,30,31] to test the standard model of
cosmology and the particle interaction theories to much better precision
level than earlier. So the precise calculations are needed to correlate
observational data with theoretical models and use data to test models; or
get help from theoretical models to develop techniques to make observations
more precise. So the first order results may not be accurate enough to meet
the precision level of future probes and we need to go to second order of
perturbation. However, all of the previously established relations between
nucleosynthesis parameters and the electron mass are acceptable
approximations for the first order corrections used in the last section.
Also FTD corrections can be studied independently at the one loop level only
where it is possible to sagregate between radiative corrections from vacuum
and FTD corrections due to the interactions with the statistical background,
in real time formalism. The validity of these expressions for the higher
loops is questionable as the overlap between the hot and cold loops increase
the contributions of hot terms and they are not totally different from cold
contributions. So we cannot calculate the second order thermal contributions
individually. They are always entangled with the cold terms and should be
put together and then, if possible, hot contributions may be separated out
at the two-loop level. When we go to check the effects of higher order
background contributions, we may need to re-write or at least re-check the
validity of these relations. Moreover, thermal contributions to electron
mass during nucleosynthesis are so complicated [19] that even to extract the
correct thermal behavior, simple analytical methods cannot be used and
numerical evaluation of hot terms will be needed. Just for comparison
betweeen one-loop and two loop behavior; we mention existing approximate
results [32].

Using second order contributions to the electron mass at low temperature $%
(T<m)$, leading order contributions to the helium yield can be computed as

\begin{equation}
\Delta Y\overset{T<m}{\longrightarrow }0.04\left( \frac{m}{T}\right)
^{2}\left( \frac{1}{3}\frac{\alpha \pi T^{2}}{m^{2}}+15\alpha ^{2}\left( 
\frac{T^{2}}{m^{2}}\right) \right) ,
\end{equation}%
whereas, the leading order contributions at high temperature $(T>m)$ comes
out to be

\begin{equation}
\Delta Y\overset{T>m}{\longrightarrow }0.04\left( \frac{m}{T}\right)
^{2}\left( \frac{1}{2}\frac{\alpha \pi T^{2}}{m^{2}}+\frac{\alpha ^{2}\pi
^{2}}{4}\left( \frac{T^{2}}{m^{2}}\right) ^{2}-\frac{\alpha ^{2}}{4}\frac{%
m^{2}}{T^{2}}\right) .
\end{equation}%
Eqs. (31)\ and (32) correspond to second order corrections, in these
approximate methods. Eq. (31) shows that the second order corrections are
sufficiently smaller than the first order corrections at low temperature
given as $3.38\times 10^{-4}$ in comparison with the one loop low
temperature contribution of $3.06\times 10^{-4}$. However, the high
temperature contributions from Eq. (32) are not very encouraging as they not
only reduce the helium abundance at high temperature, the last term on the
right hand side of equation induces very strong temperature dependence at
lower temepratures and helium yield becomes negative before the
nucleosynthesis is started. This unusual behavior has to be carefully
studied, even before we look for its physical interpretation.

\section{Results and Discussion}

Nucleosynthesis plays an important role in understanding the astrophysical
problems such as the matter creation in the early universe, inflation,
energy density of the universe and stellar structure formation. Primordial
nucleosynthesis in a very hot and extremely low density universe was
significant until the production of $^{4}He$ and has been studied in detail,
not only to resolve some key issues of SBBM of cosmology; but also some
important issues of nuclear interactions in high energy physics. Study of
nucleosynthesis also helps to understand large scale structure formation,
energy density of the universe and chemical evolution of galaxies. There are
theoretical as well as observational predictions for primordial $^{4}He$
yield. Beta decay and weak interactions played an important role during the
primordial nucleosynthesis, until it freezed out. Afterward the temperature
was lowered and the available neutrons fused to form the light nuclei. The $%
^{4}He$ abundance parameter is however sensitive to the electron mass and
the temperature dependence of phase space due to the Fermi-Dirac
distribution function for hot electrons and Bose-Einstein distribution
function for the bosons. However, the radiative corrections are not very
important as we do not have to include radiative corrections to the decay
rates [2,28], at least at the one-loop level. Therefore, the existence of
finite temperature QED background makes it relevant to include its
corrections to electron propagator. Background corrections mainly appear
from the selfmass of electron. In the previous section, we calculated FTD
corrections to different parametrs of cosmology in terms of $\frac{\delta m}{%
m}$. We list the numerical values of these parameters as a function of
temperature. Table 1 indicates the numerical values of all these parameters
and shows the difference of values of these parameters at T=m for a cooling
and a heating system

\bigskip

. It is clear from the above equations that all of nucleosynthesis
parameters become slowly varying function of temperature during
nucleosynthesis. They are constant before and after the nucleosynthesis. All
of the nucleosynthesis parameters are plotted as a function of temperature
at the one loop level. All of these parameters are constant for low and high
temperatures and become slowly varying functions of temperatures during
nucleosynthesis. The temperature and density dependence of these functions
can be expressed in terms of $a(m\beta ,\pm \mu )$, $b(m\beta ,\pm \mu )$%
\bigskip \ and $c(m\beta ,\pm \mu )$ functions [2, 16, 17, 25, 26] in the
corresponding ranges of temperature and density through $\frac{\delta m}{m}$
given in Eqs. (14, 15 and 21), up to the one loop level in perturbation
theory.

\FRAME{ftbpF}{1.7045in}{1.2358in}{0pt}{}{}{OneLoopTwoLoop.eps}{\special%
{language "Scientific Word";type "GRAPHIC";maintain-aspect-ratio
TRUE;display "USEDEF";valid_file "F";width 1.7045in;height 1.2358in;depth
0pt;original-width 10.6562in;original-height 7.7141in;cropleft "0";croptop
"1";cropright "1";cropbottom "0";filename
'OneLoopTwoLoop.eps';file-properties "XNPEU";}}Higher order contributions to
the nucleosynthesis are not so simple and cannot really be evaluated without
numerical computations which are out of scope of this paper. However, the
second order behavior is expected to be similar to the one-loop level. They
are almost constant for low temperatures. However, during and after
nucleosynthesis, their estimated approximate behavior given in Figure 5 may
not be a good approximation. It can be easily seen from the graph that
two-loop contribution is much smaller than one loop level and can easily be
ignored. However, due to a negative term in Eq.(32) the helium sythesis
seems to be decreasing rather than increasing at high temperatures. This
leading order behavior has to be studied, in detail, as it may help to let
us understand the universe in a little better way and or help to understand
the limits on the validity of the theory or the calculational scheme.
Numerical calculations of the thermal corrections at the higher loop level
will be helpful at this stage.

The data from WMAP is still being interpreted and the later observational
probes such as Planck [33] and James Webb Space Telescope [34] are expected
to provide further fine tuning in precision values of these parameters. For
this, even higher order modifications to electron mass at finite
temperatures may provide further refinement to such corrections.$\bigskip $

\section{References and Footnotes}

\begin{enumerate}
\item See for example, Utpal Sarkar, Particle and Astroparticle Physics'
(Taylor \&Francis Group 2008); and

S.Weinberg `Gravitation and Cosmology' (Wiley, New York, 1972) and
references therein.

\item Samina Saleem (Masood), Phys. Rev. D36 , 2602 (1987).

\item Samina Masood, BAPS.2013.APR.S2.11. :\textbf{arXiv:1205.2937 [hep-ph].}

\item Samina Masood,`Renormalization of QED\ Near Decoupling Temperature',
(submitted for publication)

\item Samina S.Masood,Phys.Rev.D48, 3250(1993) and references therein.

\item Samina S.Masood, Astroparticle Phys. 4, 189 (1995).

\item Samina S. Masood, et.al; Rev.Mex.Fis. 48, 501 (2002).

\item M.Chaichian, et.al; Phys.Rev.Lett. Phys. Rev. Lett. 84 5261 (2000).

\item Samina S. Masood,`Magnetic Moment of Neutrino in the strong magnetic
field' ,\textbf{\ arXiv: hep-ph/0109042. }

\item Samina S.Masood, `Scattering of Leptons in Hot and Dense Media', 
\textbf{arXiv: hep-ph/0108126.}

\item Samina S.Masood and Mahnaz Qader,Phys.Rev.D46,5110(1992).

\item K.S.Babu and V.S.Mathur, Phys.Lett. B196, 218 (1987); M.Fukujita and
T.Yanagida, Phys.Rev.Lett.58,1807(1987)

\item Pal and Mohapatra, Massive Neutrinos in Physics and Astrophysics'
(World Scientific publication, 1991)

\item L.Dolan and R.Jackiw, Phys.Rev. D9,3320(1974)

\item P.Landsman and Ch. G. Weert, Phys.Rep.145,141(1987) and references
therein

\item See for Example, K.Ahmed and Samina Saleem (Masood) Phys.Rev
D35,1861(1987).

\item K.Ahmed and Samina Saleem (Masood), Phys. Rev D35, 4020(1987).

\item K.Ahmed and Samina Saleem (Masood),Ann.Phys.164,460(1991) and
references therein.

\item E.Braaten and R.D.Pisarski, Nucl. Phys. B339,310(1990); ibid, B337,
569(1990)

\item Mahnaz Qader, Samina S.Masood and K.Ahmed, Phys.Rev.D44,3322(1991).
Mahnaz Qader, Samina S.Masood and K.Ahmed, D46,5633(1992)and references
therein.

\item Samina Masood and Mahnaz Haseeb, Int.J.Mod.Phys. A27 (2012) 1250188;

\item Samina Masood and Mahnaz Haseeb, Int.J.Mod.Phys. A23 (2008) 4709-4719.

\item Mahnaz Haseeb and Samina Masood, Chin.Phys. C35 (2011) 608-611;

\item Mahnaz Haseeb and Samina Masood, Phys.Lett. B704 (2011) 66-73 and
references therein.

\item R.Kobes and G.W.Semenoff, Nucl.Phys. B260,714(1985); ibid,
B272,329(1986)

\item See for example,L.D.Landau and E.M.Lifshitz,`Statistical
Physics'(Pergamon Press L.D.LandauLondon,1968)

\item E.J.Levinson and D.H.Boal, Phys.Rev. D31,3280(1985).

\item Samina S.Masood, Phys.Rev.D44,3943(1991).

\item Samina S.Masood,,D47,648(1993) and references therein.

\item See for Example, A.Weldon, Phys.Rev.D26,1394(1992); ibid;2789(1982).

\item J.L.Cambier, J.R.Primack and M.Sher, Nucl. Phys., B209, 372 (1982);
ibid B222, 517E (1983); D.A.Dicus, E.W.Kolb, A.M.Gleeson, E.C.G.Sudershan,
V.L.Teplitz and M.S.Turner, Phys. Rev., D26, 2694 (1982).

\item ALEPH Collaboration, Phys.Lett. B499 (2001) 53-66.

\item D. N. Spergel et. al, Wilkinson Microwave Anisotropy Probe (WMAP)
Three Year Observations: Implications for Cosmology, Astrophys. J. Suppl.
170 (2007) 377-408.

\item Jens Chluba et, al.,Probing the inflaton: Small-scale power spectrum
constraints from measurements of the CMB energy spectrum Astrophys.J. 758
(2012) 76.

\item Mahnaz GHQ.Haseeb, Obaidullah Jan and Omair Sarfaraz, arXiv:1111.0248v2

\item %
http://www.nature.com/news/planck-telescope-peers-into-primordial-universe-1.12658

\item http://www.jwst.nasa.gov/
\end{enumerate}

\end{document}